\documentstyle[12pt,epsfig]{article}
\input epsf

\def\frac#1#2{{#1\over #2}}
\def\lsim{\mathrel{\rlap{\lower4pt\hbox{\hskip1pt$\sim$}}
    \raise1pt\hbox{$<$}}}         %less than or approx. symbol
\def\gsim{\mathrel{\rlap{\lower4pt\hbox{\hskip1pt$\sim$}}
    \raise1pt\hbox{$>$}}}         %greater than or approx. symbol

\def\pom {\hbox{I$\!$P }}
\def\odd {\hbox{$\!$O }}
\def\t{$|t|$ }

\begin{document}

\rightline{LYCEN 9895 (November 1998)}

\bigskip
\bigskip
\begin{center}                                                                 
{\Large \bf OF DIPS AND STRUCTURES}
\bigskip 
\bigskip

{\bf P. Desgrolard}({\footnote{E-mail: desgrolard@ipnl.in2p3.fr}}),
{\bf M. Giffon}({\footnote{E-mail: giffon@ipnl.in2p3.fr}}),
{\bf E. Martynov}({\footnote{E-mail: martynov@bitp.kiev.ua}}),
{\bf E. Predazzi}({\footnote{ E-mail: predazzi@to.infn.it}}).

\bigskip
($^{1,2}$){\it 
Institut de Physique Nucl\'eaire de Lyon, IN2P3-CNRS et Universit\'{e}
Claude Bernard,\\
43 boulevard du 11 novembre 1918, F-69622 Villeurbanne Cedex, France\\}

($^3$){\it
N.N. Bogoliubov Institute for Theoretical Physics, National Academy of
Sciences of Ukraine,
252143, Kiev-143, Metrologicheskaja 14b, Ukraine\\}

($^4$){\it                                                                  
Dipartimento di Fisica Teorica - Universit\`a di Torino 
and Sezione INFN di Torino, Italy}

\end{center}
                                                                               
\bigskip
\bigskip

{\bf Summary}
A detailed analysis of the existence of high energy
secondary diffractive dips and 
structures in the extrapolations of the fits to the data is given.
The existence of these dips and {\it a fortiori} their 
position is found to be rather model-dependent~: present 
in all eikonalized models including Pomeron, Odderon and
secondary Reggeons they disappear when an additional large-\ t term is added (as sometimes advocated). 

%\end{titlepage}

\bigskip
\bigskip

{\Large \bf 1. Layout of the paper}

\medskip

 Few years ago, the suggestion was made \cite{dgp} that, at increasingly
high
energies, secondary (diffractive) dips and structures should develop at
intermediate $|t|-$values in both
 $pp$ and $p\bar p$ angular distributions. In
particular, in \cite{dgp} it was suggested that such effects should be well
visible at LHC while only extremely precise data could perhaps show up 
 at RHIC energies. As a matter of fact, predictions of secondary
structures have
 appeared many times in the past \cite{past}.
  The large spectrum of predictions in the position of these
secondary dips shows that things are actually more complicated than anticipated
long ago \cite{zac}. It is not enough that a given scheme inherently generates
oscillations (like the Bessel function of an impact parameter 
representation); interference effects are very important in determining
 their
position. The model dependence 
 of these predictions, however, is
 not so important; it is the prediction itself of the existence of
secondary structures which matters. This was received with great interest and
plans are under the 
way to investigate this point experimentally \cite{exp}. 
In \cite{dgp} the prediction was based on the extrapolation to higher energies
of high quality fits to all existing data and proposed {\it "...with the greatest
reservation..."} since it was the result of dealing with the data within fairly
sophisticated "Born amplitudes" devised to provide the best possible fit. The
question which was left open to further investigation was how credible this
extrapolation could be and how much it could be generalized. For this reason,
a careful analysis of different models has now 
been carried out
using various refinements. In particular, we have undertaken an extensive
analysis of several models where the input (or Born amplitude) is variously
eikonalized as a sort of unitarization \cite{uni}.

As a result of these investigations, in the present paper we reconsider the
entire question by~:
{\it (i)} using more general schemes of eikonalization
\cite{uni,uni2,dgmp};
{\it (ii)} choosing various kinds of input amplitudes (in particular, for their
simplicity, monopoles \cite {mon} and dipoles \cite{las,dgj}, see below);
{\it (iii)} occasionally superimposed to a "large-$|t|$ term" attributed \cite{dl}to 
Odderon exchange and behaving 
like $t^{-4}$ \cite{dl} at hight \t.
 
Here, we will not report all the results obtained so far but we will rather 
give a general summary moving quickly to the
conclusions (a more complete analysis and discussion will be reported soon 
\cite{dgmp}).

At high (LHC) energies, if we strictly respect the unitarity constraints
\footnote{The data at the present highest available energies are, notoriously,
still far from asymptopia (and this appears quite clearly from our analysis 
\cite{dgmp}). It might even be that it will never be reached in actual experiments.
In this sense it will be necessary to check at each step that unitarity is not
violated; as an example, it is not sufficient to simply state that {\it Froissart's 
bound is not violated}.}
and confining ourselves to just one case (chosen as an example to be a dipole
\cite{las,dgj} for both Pomeron and Odderon Born amplitudes plus secondary
Reggeons) our conclusions are~:

(a) secondary diffractive structures are invariably predicted after
eikonalization of the Born amplitude (so long as a large-$|t|$ term is
not included).

(b) The position of the dips is quite model dependent. Loosely speaking, the
first se\-condary dip at $\sqrt s= 14$ TeV is predicted around $|t_2|\sim 1.5-3$
GeV$^2$. 

(c) When a large-$|t|$ term is altogether omitted, 
the differential cross section is dominated by the eikonalized Odderon term 
as $|t|$ increases (this term being quite negligible at small \t values).

(d) All secondary structures tend to disappear if a large-$|t|$ term is 
surimposed to the previous amplitude. Possiby, this amounts to double counting 
but this point is far from clear.
 
At lower (RHIC) energies, the situation appears even more problematic because 
we find, at best, some hint of new structures which manifest themelves as breaks
of the slopes.

\bigskip

{\Large \bf 2. The input Born }

\medskip

We focuse on the (dimensionless) crossing-even and -odd amplitudes $a_\pm(s,t)$ 
of the $pp$ and $\bar pp$ reactions 
$$A_{pp, {\rm Born}}^{\bar pp}(s,t)=a_+(s,t)\ \pm a_-(s,t)\ ,
\eqno(1)$$
for which we have data on

\noindent
i) the total cross-sections
$$\sigma_t = {4\pi\over s}\Im{\rm m} A(s,t=0) \ ,                   \eqno (2)$$
ii) the differential cross-sections
$${d\sigma \over dt}={\pi\over s^2}\big|A(s,t)\big|^2 \ ,           \eqno (3)$$
iii) the ratio of the real to the imaginary forward amplitudes
$$\rho={\Re{\rm e} A(s,t=0)\over \Im{\rm m} A(s,t=0)} \ .            \eqno(4)$$
The crossing even part in the Born amplitude is a Pomeron (to which an
$f-$Reggeon is added) 
while the crossing odd part is an Odderon 
(plus an $\omega-$Reggeon)
$$ a_+(s,t)=\ a_{\pom} (s,t)+\ a_f(s,t) \ ,\quad
   a_-(s,t)=\ a_{\odd} (s,t)+\ a_\omega(s,t) \ .                     \eqno(5)$$
For simplicity the two Reggeons that have been
retained have been taken in the standard form (with fixed 
parameters \cite{dgp} for economy)
$$ a_R(s,t)= a_R\tilde s^{\alpha_R(t)}\ ,
\quad \alpha_R(t)=\alpha_R(0) + \alpha'_R t \ ,
\quad (R=f\, {\rm and}\, \omega)  \ ,                                \eqno(6)$$
where $a_f$ ($a_\omega$) is real (imaginary).

For the Pomeron $a_{\pom}(s,t)$ and the Odderon 
$a_{\odd}(s,t)$, in this paper we take just                         
a dipole $D(s,t)$  (a double pole in the complex angular momentum) of the form 
\footnote{In (7) and (9), a suffix \pom or \odd is
understood according to wether we are referring to the Pomeron or to the
Odderon. With our choices, $a_{\pom}$ is real and $a_{\odd}$ is imaginary}
$$ D(s,t)= a \ \tilde s^{\alpha(t)}\left[e^{b(\alpha(t)-1)} (b+\ell n{\tilde s})
                                   \ +\ d \ell n{\tilde s}\right] \ .\eqno(7)$$
As usual  
$$\tilde s\ =\ {s\over s_0} \ e^{-i{\pi\over 2}}\ ,\quad 
(s_0=1\ {\rm GeV}^2)                                                   \eqno(8)$$ 
(to respect $s-u$ crossing)
and $\alpha(t)$ is the Regge trajectory taken of the linear form\footnote{Linear trajectories are an oversimplification that, strictly, violates
analyticity. In addition, at large-\t this may be dangerous in practice. We 
ignore this complication.}    
$$\alpha(t)= 1+\delta+\alpha't  \  .                                \eqno(9)$$

The case where the input is a monopole (i.e. a simple pole in the angular momentum plane) will be considered in 
\cite{dgmp} (the difference  between a monopole and a dipole is
essentially that the amplitude for the second grows with an additional power of
$\ell n s$). It is quite difficult to discriminate between these forms on general
grounds as well as phenomenologically since a reasonable or good agreement
with the data is obtained with both and the reason behind this is, presumably,
that the data are not yet asymptotic as already mentioned.
 
Some authors maintain that a perturbative large-\t term behaving like $|t|^{-4}$
(and complying with perturbative QCD requirements according to \cite{dl} is to
be added to the Odderon\footnote{We should, however, not forget that at
large-\t, the ratio $|t|/s$ is really rather small so that we are in a domain
closer to the usual Regge kinematics than to that of perturbative QCD.}.  We
believe that, when the Born amplitude is eikonalized, the rescattering
corrections implied by eikonalization should be the end of story especially for
trajectories rising slower than linearly.  Adding another term could lead to
double counting.  In \cite{dgmp} however, we will investigate for completness
the r\^ole of incorporating in the Odderon an additional large \t term.

\bigskip

{\Large \bf 3. Eikonalization procedure}

\medskip

A positive $\delta$ value (for either \pom or \odd or both) in (9)
signals a supercritical situation necessitating some kind of regularization 
(like eikonalization) to avoid conflicts with unitarity.
Several eikonalization procedures can be found in the litterature (see 
\cite{uni,uni2,dgmp} and references therein). They all amount to taking 
rescattering (therefore, hopefully, unitarity) corrections into account.
If $a_\pm(s,t)$ are our Born amplitudes (5),
the crossing even and crossing odd input amplitudes in the impact
parameter or $b$- representation are proportional to the {\it eikonal
function} $\chi_\pm (s,b)$
$$h_\pm \equiv h_\pm(s,b)=\ {1\over 2}\chi_\pm (s,b)=\
 {1\over 2s}\int_0^\infty \ dq \ q\, J_0(bq)\, a_\pm
(s,-q^2)\  ,\quad                  (q^2=-t) \ .                   \eqno(10)$$
 
In a generalized scheme that accounts for rescattering corrections,
one can prove (see \cite{dgmp} for details) that the impact 
parameter amplitude takes the two parameter form
\footnote{This procedure, roughly speaking, mimicks a situation whereby 
the particle-Pomeron-particle
and the particle-Odderon-particle amplitude vertices ($g_+$ and $g_-$) are 
rescaled by {\it a priori} different positive constants 
$(\sqrt{\lambda_+}$ and $\sqrt{\lambda_-})$.}
$$H^{\bar pp}_{pp}(s,b)\ =\ h_+\pm h_-$$
$$+\ \left(\frac{h_+\sqrt{\lambda_+} \pm h_-\sqrt{\lambda_-}}
{h_+\lambda_+\pm h_-\lambda_-}\right)^2
\left(\frac{e^{2i(h_+\lambda_+\pm h_-\lambda_-)}-1}{2i}
- (h_+{\lambda_+}\pm h_-{\lambda_-})               \right)\       .\eqno(11)$$

This corresponds to the so-called {\it generalized eikonalization} \cite{uni}.
When $\lambda_+=\lambda_-$, (11) reduces to the {\it quasi-eikonalization}
\cite{uni2}.  This reduces further to the special (and traditional) eikonal
form\footnote{ We do not expect this standard eikonalization process to give
satisfactory results with the simple Born models of Sect.2; this is why a
generalization is considered.}

$$H^{\bar pp}_{pp}(s,b)\ =\frac{e^{2i(h_+\pm h_-)}-1}{2i}\
,                                                                  \eqno(12)$$ 
when $\lambda_+=\lambda_-=1$ .
  
Once the eikonal amplitude $H^{\bar pp}_{pp}(s,b)$ is known in the $b-$representation, the inverse Fourier-Bessel transform leads, finally, to the  
eikonalized amplitude 
$$A^{\bar pp}_{pp, {\rm Eik}}(s,t)=\ 2s \int_0^\infty \ db \ b\, 
J_0(b\sqrt{-t})\, H^{\bar pp}_{pp}(s,b)\  .                        \eqno(13)$$

The consequences of the unitarity constraint
$\left| H(s,b)\right| \le 1 \ {\rm and }  \,\, \Im {\rm m}H(s,b)\ge 0$

$$\alpha'_{\pom} \ge \alpha'_{\odd}, \delta_{\pom} \ge \delta_{\odd}
                                                                  \eqno(14)$$
and other limitations are discussed in great details elsewhere \cite{fm}. The r\^ole of 
unitarity and its constraints on (11) will be discussed at length in
\cite{dgmp}). Here, however, they are carefully taken into account.

\bigskip

{\Large \bf 4. (Some) results and conclusions.}

\medskip

As already stated, in this paper we consider only the dipole case
without a large-$|t|$ term added to the Odderon and we compare the full case of
(11) ($\lambda_+ \ne \lambda_-$) with the one for which $ \lambda_+ =
\lambda_-$. A good reproduction of the data is obtained where the most
relevant parameters are $\lambda_+ =0.5;\ \lambda_- =1.32;\ \delta_{\pom} = 
0.058;\ \delta_{\odd} \simeq 0$ for the first case 
and $\lambda = 0.5;\ \delta_{\pom} = 0.068;\ \delta_{\odd}=0$ for the second one.

The extrapolation to 500 GeV and 14 TeV, (the energies to be reached in the near
future \cite{exp}), are shown in Fig.1. We find

(a) for the first fit ($\lambda_+ \ne \lambda_-$ ) $\chi_{/dof}^2=11.1$;
 
(b) for the second one ($\lambda_+ = \lambda_-$) $\chi_{/dof}^2=14.1$. 

\smallskip
Several comments are in order~:

Secondary structures are present in both cases ($\lambda_+\neq \lambda_-$ and $\lambda_+=\lambda_-$) and
a second dip clearly appears as the energy increases.
However, from an extensive trial of all possibilities (not reported here), we
find that its location moves considerably from one case to the other.
If by fault of better argument, we had to
choose the best fit, the high energy extrapolation (LHC) would favor 
$|t_2|\simeq 2.7$ GeV$^2$ for the case where $\lambda_+ \ne \lambda_-$ and  
$|t_2|\simeq 1.7$ GeV$^2$ for the case where $\lambda_+ = \lambda_- $.

Nothing really conclusive we feel may be said at RHIC energy where, at best, a
break in the slope can be seen.

The addition of a large-\t term as advocated in \cite{dl} leads in both cases to
fits qualitatively good and comparable to those which may be obtained at the
Born level \cite{dgp}.  As already stated, in both cases, if a large-\t term is
added, the secondary structures disappear from our predictions.  We will discuss
this in \cite{dgmp} but we really do not know why this happens; the best we can
offer is that adding a large-\t term to the eikonalized Odderon tautamounts to
double counting and destrys the simplicity of the approach (see also comments at
the closing of Section 2).

We conclude that a careful reanalysis based on full eikonalization confirms our
early prediction about the existence of secondary diffractive structures 
in the case of a crude dipole model unless a large-\t term is added.
 
Searching for the origin of secondary structures we find that this is a
cooperative effect due to the eikonalized Pomeron and Odderon.  If one removes
all components one by one, when we are left with a dipole Pomeron alone,
structures are present but the fit to the large-\t data becomes disastrous.  The
large-\t domain is dominated by the eikonalized Odderon and this alone shows
also structures (but the fit is {\it a fortiori} disastrous).  Keeping both
eikonalized Pomeron and Odderon (no secondary Reggeons) reproduces roughly the
data and shows secondary structures.

Based on a wide exploration of simple models, we feel that we can commit
ourselves to saying that the second minimum at the LHC may be expected anywhere
between

$$ 1.5\le |t_2|\ {\rm (GeV}^2)\le 3 .   \eqno(15) $$

Only the experiment can answer the question of the existence and location of
secondary dips and discriminate among models.

\begin{center}
\leavevmode\epsfxsize=14cm\epsfbox{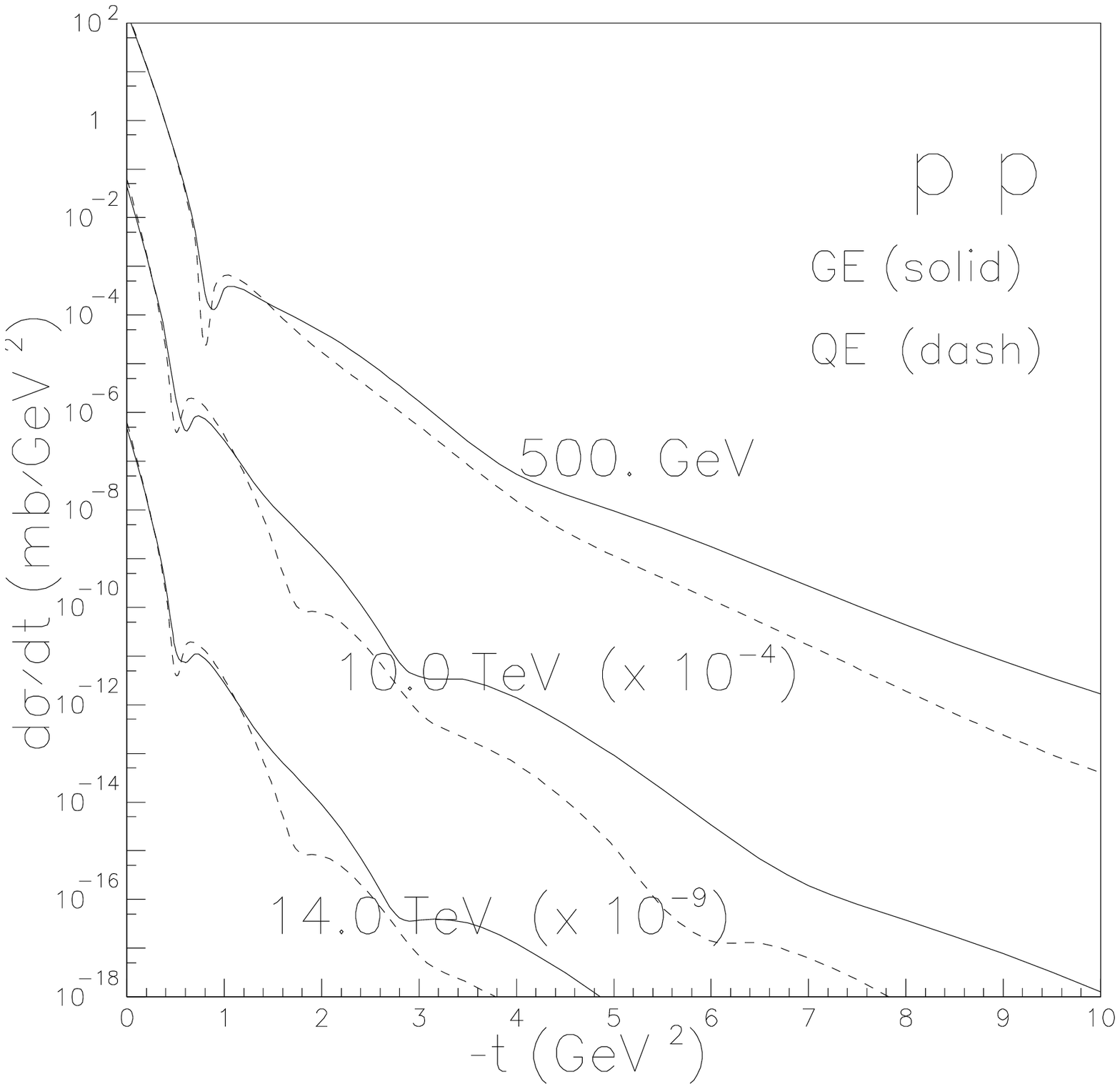}
\end{center}

\bigskip
{\bf Fig. 1} Extrapolations to RHIC and LHC energies of the angular 
distributions for two fits using the dipole model (dipole Pomeron, dipole 
Odderon and Reggeons) respecting the unitarity constraints~: 
(a) dash lines~: quasi-eikonalization procedure (QE).
(b) solid lines~: generalized eikonalization procedure (GE).


\begin{thebibliography}{99}
\bibitem{dgp} P. Desgrolard, M. Giffon and E. Predazzi, Zeit. Phys. C {\bf
63} (1994) 241.                                                             
\bibitem{past} Earlier predictions of secondary structures are~:

{ (a)} T.T. Chou and C.N. Yang, in  "High Energy Physics and Nuclear
Structure", edited by G. Alexander (Horth Holland, Amsterdam) 1967, p.348;
T.T. Chou and C.N. Yang, Phys. Rev. {\bf 170} (1968) 1591;
T.T. Chou and C.N. Yang, Phys. Rev. Lett. {\bf 20} (1968) 1213;
T.T. Chou and C.N. Yang, Phys. Rev. {\bf D17} (1978) 1889.

{ (b)} L.Durand III and R.Lipes, Phys. Rev. Lett.{\bf 20} (1968) 637.

{ (c)} C. Bourrely, J. Soffer and T.T. Wu, Phys. Rev. {\bf D19} (1979)
3249;
C. Bourrely, J. Soffer and T.T. Wu, Nucl. Phys. {\bf B247} (1984) 15;
C. Bourrely,
in {\it Elastic and Diffractive Scattering at the Collider and beyond}, 
1$^{\rm st}$ International Conference on Elastic
and Diffractive Scattering, Chateau de Blois, France (1$^{\rm st}$
"Blois workshop")- June 1985, edited  by B. Nicolescu  and J. Tran Thanh Van
(Editions Fronti\`eres), p. 239.

{ (d)} M.J. Menon, Nucl. Phys. {\bf B} (Proc. Suppl.) 25B (1992) 94, and
references therein.

\bibitem{zac} D. Horn and F. Zachariasen in "Hadron Physics at very high Energies"
(W.A. Benjamin {\it inc}., 1973). 

\bibitem{exp}
 (a) W. Guryn {\it et al.}, in {\it Frontiers in Strong Interactions},
VII$^{\rm th}$ Blois Workshop on
Elastic and Diffractive Scattering, Chateau de Blois, France - June 1995,
edited by P. Chiappetta, M. Haguenauer and J. Tran Thanh Van (Editions
Fronti\`eres) 1996, p. 419.

(b) M. Buenerd {\it et al.}, in {\it Frontiers in Strong Interactions},
VII$^{\rm th}$ Blois Workshop on Elastic
and Diffractive Scattering, Chateau de Blois, France - June 1995, edited by P.
Chiappetta, M. Haguenauer and J. Tran Thanh Van (Editions Fronti\`eres) 1996,
p. 437; The TOTEM Collaboration,
{\it Total Cross-section, Elastic Scattering and Diffraction
Dissociation at the LHC}, CERN/LHC 97-49 LHCC/I 11 (August 1997).

\bibitem{uni} For a summary on the relationship between eikonalization and
unitarization, see
M. Giffon, E. Martynov and E. Predazzi, Zeit. Phys. C {\bf 76} (1997) 155.

\bibitem{uni2} For such general eikonalization schemes, see 
K. A. Ter-Martirosyan, Sov. ZhETF Pisma {\bf 15} (1972) 519;
A. Capella, J. Kaplan and J. Tran Thanh Van, Nucl. Phys. {\bf B97} (1975) 493; 
A. B. Kaidalov, L. A. Ponomarev and K. A. Ter-Martirosyan,
Sov. Journ. Part. Nucl. {\bf 44} (1986) 468; S.M. Troshin and N.E. Tyurin, Sov. J.
Part. Nucl. {\bf 15} (1984) 25.
 
\bibitem{dgmp} Further work will be reported soon~: P. Desgrolard, M. Giffon, 
E. Martynov, E. Predazzi (in preparation).

\bibitem{mon} P.D.B. Collins, in {\it An Introduction to Regge Theory and
High-Energy Physics}, (Cambridge University Press, Cambridge) 1977.

\bibitem{las} L. Jenkovszky, Fortsch. Phys. {\bf 34} (1986) 702;
L. Jenkovszky, A.N. Shelkovenko and B.V. Struminsky, Zeit. Phys. C {\bf 36} (1987)
495.

\bibitem{dgj} P. Desgrolard, M. Giffon and L. Jenkovszky,
Zeit. Phys. C {\bf 55} (1992) 637.

\bibitem{dl} A. Donnachie and P.V. Landshoff, Zeit. Phys. C {\bf 2} (1977) 55;
Nucl. Phys. {\bf B 231} (1984) 189.

\bibitem{fm} J. Finkelstein, H.M. Fried, K. Kang, C.-I. Tan, Phys. Lett. B {\bf 232} (1989) 257;
E.S. Martynov,  Phys. Lett. B {\bf 232} (1989) 367.
\end{thebibliography}
\end{document}